 \documentclass[smallabstract,smallcaptions]{dccpaper}

\usepackage{epsfig}
\usepackage{amsmath}
\usepackage{amssymb}
\usepackage{color}
\usepackage{url}
\usepackage{amsfonts}
\usepackage{graphicx}
\usepackage{hyperref}
\usepackage{booktabs}

\newlength{\figurewidth}
\newlength{\smallfigurewidth}

\begin{document}

\title
{\large
\textbf{Machine Perceptual Quality: Evaluating the Impact of Severe Lossy Compression on Audio and Image Models}
}

\author{%
Dan Jacobellis$^{\ast}$, Daniel Cummings$^{\dag}$, and Neeraja J. Yadwadkar$^{\ast}$\\[0.5em]
{\small\begin{minipage}{\linewidth}\begin{center}
\begin{tabular}{ccc}
$^{\ast}$University of Texas at Austin & \hspace*{0.5in} & $^{\dag}$Intel Labs, Intel Corporation \\
Austin, TX, 78712, USA && Austin, TX, 78746, USA\\
\url{danjacobellis@utexas.edu} && \url{daniel.cummings@intel.com} \\
\url{neeraja@austin.utexas.edu} 
\end{tabular}
\end{center}\end{minipage}}
}

\maketitle
\thispagestyle{empty}

\begin{abstract}
In the field of neural data compression, the prevailing focus has been on optimizing algorithms for either classical distortion metrics, such as PSNR or SSIM, or human perceptual quality. With increasing amounts of data consumed by machines rather than humans, a new paradigm of machine-oriented compression---which prioritizes the retention of features salient for machine perception over traditional human-centric criteria---has emerged, creating several new challenges to the development, evaluation, and deployment of systems utilizing lossy compression. In particular, it is unclear how different approaches to lossy compression will affect the performance of downstream machine perception tasks. To address this under-explored area, we evaluate various perception models—including image classification, image segmentation, speech recognition, and music source separation—under severe lossy compression. We utilize several popular codecs spanning conventional, neural, and generative compression architectures. Our results indicate three key findings: (1) using generative compression, it is feasible to leverage highly compressed data while incurring a negligible impact on machine perceptual quality; (2) machine perceptual quality correlates strongly with deep similarity metrics, indicating a crucial role of these metrics in the development of machine-oriented codecs; and (3) using lossy compressed datasets, (e.g. ImageNet) for pre-training can lead to counter-intuitive scenarios where lossy compression increases machine perceptual quality rather than degrading it. To encourage engagement on this growing area of research, our code and experiments are available at: \url{https://github.com/danjacobellis/MPQ}.

\end{abstract}

\begin{figure}[t]
\begin{center}
\epsfig{width=6.0in,file=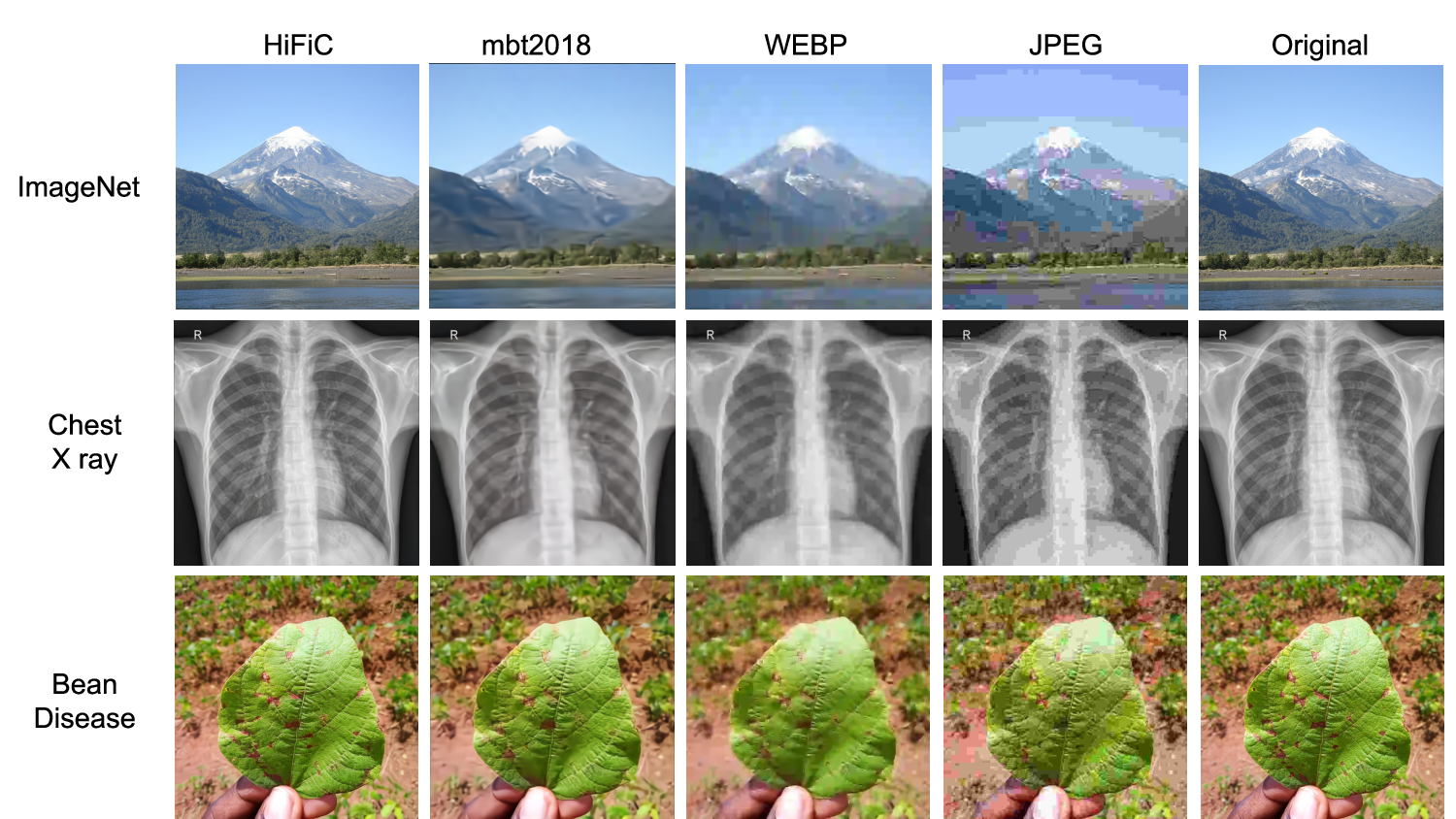}
\end{center}
\caption{\label{fig:image_compression_methods}%
Visual comparison of image compression methods. The original ImageNet image is JPEG compressed at near-lossless quality level of 96 (5.1 BPP), while the Chest X-ray and bean disease original images are lossless.}
\end{figure}

\Section{Introduction}

In contemporary machine perception pipelines, lossy compression techniques are often employed, but using legacy codecs at near-lossless quality levels, thus limiting potential savings in data rate~\cite{ehrlich2022first}. For instance, the ImageNet dataset, a cornerstone for image classification tasks, utilizes JPEG compression with an average compression ratio of roughly 5:1. As a result, the full ImageNet-21k is over 1.3 TB in size, and common practice is to discard most of this information using a 224x224 reduced resolution version~\cite{ridnik2021imagenet}. Additionally, many types of sensors necessitate extremely high compression ratios, sometimes exceeding 1000:1~\cite{cocker2022low}, resulting from high resolution measurements combined with limited communication bandwidth. As a result, vast amounts of rich, high-fidelity data captured by modern sensors are underutilized or even discarded entirely.

In the decades since the introduction of the ubiquitous JPEG and MPEG standards for images and audio, advancements in lossy compression technologies have demonstrated the capability to achieve high compression ratios with minimal degradation in quality. For example, it has been shown that storing the ImageNet-1k dataset using the tokens produced by a ViT-VQGAN neural compression model saves a factor of 100:1 in storage and leads to faster and simplified training ~\cite{yu2021vector}\cite{park2023storage}. 
While the advantages of employing more potent lossy compression techniques are evident, uncertainty surrounding their impact on downstream machine perception tasks remains a significant barrier. For example, Ilyas et al.,~\cite{ilyas2019adversarial} demonstrate the existence of signal components, called non-robust features, which are highly predictive yet imperceptible to humans. Lossy compression during training is likely to eliminate these features and thereby lead to sub-optimal models~\cite{aydemir2018effects}. Additionally, failure to match the exact lossy compression method and settings during training and inference could lead to distribution shift and unpredictable model behavior.

Our work aims to systematically evaluate the impact of various types of lossy compression---both conventional and neural---on both audio and visual machine learning tasks. By understanding these effects, we aim to bridge the gap between the promising capabilities of advanced lossy compression techniques and their practical implementation in machine learning pipelines.

\Section{Background}

Conventional media compression standards (called codecs) rely on simple but effective linear transforms that exploit the redundancies of natural signals. For example, the discrete cosine transform (DCT) used in JPEG and MP3 compresses signal energy into fewer coefficients. Carefully designed quantization matrices then assign more bits to perceptually important temporal or spatial sub-bands based on models of human sensitivity. These compression techniques have remained popular for decades since they offer a decent compression rate without excessively compromising signal quality. 

Two key developments led to a greater focus on neural network based compression. Ballé et al. \cite{balle2017end} showed that autoencoders optimized end-to-end for both rate and distortion (a.k.a. rate distortion autoencoders or RDAEs) compress images more effectively than traditional codecs. In parallel, Van den Oord ~\cite{van2017neural} introduced the vector quantized variational autoencoder (VQ-VAE) as a method of representation learning. Variants of these architectures emerged specializing them for better human perceptual quality, both for audio ~\cite{zeghidour2021soundstream} \cite{defossez2022high} and images ~\cite{he2022po}. The advent of generative compression methods ~\cite{mentzer2020high} led to observation of a rate-distortion-perception trade-off~\cite{wagner2022rate}. 

In addition to better rate-distortion performance, ongoing codec development efforts also aim to optimize for machine perception, a paradigm referred to as ``compression for machines~\cite{chamain2021end}" or ``machine-oriented compression~\cite{kang2023super}." Most notably, The JPEG AI standard \cite{ascenso2023jpeg} proposes a single stream image encoder supporting multiple decoders for both human and machine perception. Harell et al.,~\cite{harell2023rate} proposed a taxonomy of three different machine-oriented compression approaches. Notable to our work is the method of full-input machine-oriented compression, where the signal is fully decoded before performing downstream tasks; this can either be achieved using an existing codec or by optimizing the compression system for the downstream task.

While the evaluation of human perceptual quality has been extensively studied, it is less clear how different types lossy compression affect machine perception. Hendrycks et al.,~\cite{hendrycks2019benchmarking} study the impact of various corruptions, including JPEG compression, on image classification performance. Matsuraba et al.,~\cite{matsubara2023sc2} study the impact of various image compression methods on classification and segmentation. Despite these contributions, a dedicated analysis of severe lossy compression effects across a variety of applications, including generative compression methods and audio models, remains unexplored and is the focus of our investigation.

\Section{Methodology}

We investigate the impact of different audio and image compression techniques on machine perceptual quality under severe lossy compression---which we define as ratios compression ratios between 20:1 and 1000:1---and compare against a baseline that does not have additional compression. We employ six datasets, seven different lossy compression methods, and use popular pre-trained models for various discriminative tasks as summarized in Tables \ref{tab:datasets_models} and \ref{tab:compression_methods}. We use the performance on the validation split of each dataset as a measure of machine perceptual quality. We evaluate the compression performance based on bitrate, conventional distortion metrics, and deep similarity metrics.

\begin{table}[ht]
\centering
\caption{Summary of Datasets and Models}
\label{tab:datasets_models}
\begin{tabular}{llll}
\toprule
Dataset & Task Type & Model & Metric \\
\midrule
ImageNet-1k & Image classification & ViT &  Top-1 Accuracy \\
ChestX-ray8 & Image classification & ViT & Top-1 Accuracy \\
Bean Disease & Image classification & ViT & Top-1 Accuracy \\
ADE20k & Image segmentation & SegFormer & Mean intersection over union \\
Common Voice & Speech recognition & Whisper & Word recognition accuracy \\
MUSDB-HQ & Music separation & Demucs v3 & Signal-to-distortion ratio \\
\bottomrule
\end{tabular}
\end{table}

\begin{table}[ht]
\centering
\caption{Summary of Compression Methods}
\label{tab:compression_methods}
\begin{tabular}{lll}
\toprule
Method & Description & Setting \\
\midrule
JPEG & Legacy transform-coding based image codec& Quality: 5 \\
WEBP & Modern transform-coding based image codec & Quality: 0 \\
MBT2018 & Neural image codec with MSE objective & Quality: 1 \\
HiFiC & Neural image codec with adversarial objective& Quality: Low \\
MP3 & Legacy transform-coding based audio codec & Bitrate: 8 kbps \\
Opus & Modern transform-coding based audio codec & Bitrate: 6 kbps \\
EnCodec & Neural audio codec with adversarial objective & Bitrate: 6 kbps \\
\bottomrule
\end{tabular}
\end{table}

\vspace{-5mm}
\paragraph{Models and datasets.} We employ the ImageNet-1k dataset for image classification, using a vision transformer (ViT) pre-trained on ImageNet-21k ~\cite{dosovitskiy2020image}. The NIH ChestX-ray8 dataset ~\cite{wang2017chestx} for pneumonia classification and the bean disease dataset \cite{singh2023classification} are also used in conjunction with an ImageNet-21k pre-trained ViT. Semantic segmentation is performed on the ADE20k dataset \cite{zhou2017scene} using the SegFormer model \cite{xie2021segformer}. The Common Voice 11.0 dataset \cite{ardila2020common} and the Whisper model \cite{radford2023robust} are used for speech recognition. Finally, the MUSDB-HQ dataset, an uncompressed version of MUSDB18 \cite{rafii2017musdb18}, and the Demucs v3 model \cite{defossez2021hybrid} are used for music source separation. These datasets and corresponding models are summarized in Table \ref{tab:datasets_models}.

\vspace{-5mm}
\paragraph{{Compression methods}}The image compression methods in our study include JPEG, WEBP, the distortion-optimized neural compression approach from Minnen et al. ~\cite{minnen2018joint} , and the generative compression method HiFiC ~\cite{mentzer2020high}. For audio compression, we use MPEG Layer III (MP3), Opus, and the neural audio model EnCodec \cite{defossez2022high}. Table \ref{tab:compression_methods} summarizes these methods. Additional implementation details and a listing of specific model variants are available in our code repository \footnote{\href{https://github.com/danjacobellis/MPQ}{Github: danjacobellis/MPQ}}.

\vspace{-5mm}
\paragraph{Evaluation metrics.} We use conventional rate-distortion metrics as well as deep similarity metrics---quality metrics derived from deep neural networks and trained to predict human judgments of quality. Each metric is calculated on a per-sample basis.
\begin{itemize}

\item \textit{Bits Per Pixel (BPP)} and \textit{Bits Per Sample (BPS)} are used to measure the rate of images and audio signals respectively. For the EnCodec model, which supports a default mode where the VQVAE codes are directly stored and a secondary mode that uses additional entropy coding, we use the default mode without entropy coding and calculate BPS using the the product of the codebook size and the number codes. For all other codecs, we directly measure the rate based on the size of the encoded file.

\item \textit{Peak Signal-to-Noise Ratio (PSNR)} is used as a conventional distortion metric for both images and audio. We represent image signals using the range $[0,255]$ and represent audio signals using the range $[-1,1]$, so \( \text{PSNR} = 20\log_{10}(255) - 10\log_{10}(\text{MSE}) \) for images and \( \text{PSNR} = -10\log_{10}(\text{MSE}) \) for audio.

\item \textit{Learned Perceptual Image Patch Similarity (LPIPS)}\cite{zhang2018unreasonable} is a deep similarity metric specifically designed for images. It captures complex perceptual differences that simpler metrics like PSNR or SSIM are insensitive to. In the table, we report \( -10\log_{10}(\text{LPIPS similarity}) \) to align it with the other quality metrics.

\item  \textit{Contrastive Deep Perceptual Audio Similarity Metric (CDPAM)} \cite{manocha2021cdpam} is a deep similarity metric is designed for audio. Like LPIPS for images, it captures perceptual differences more effectively than PSNR. Similar to LPIPS, we report \( -10\log_{10}(\text{CDPAM similarity}) \).
\end{itemize}

\Section{Results}

\begin{figure}[tb]
\begin{center}
\epsfig{width=6.0in,file=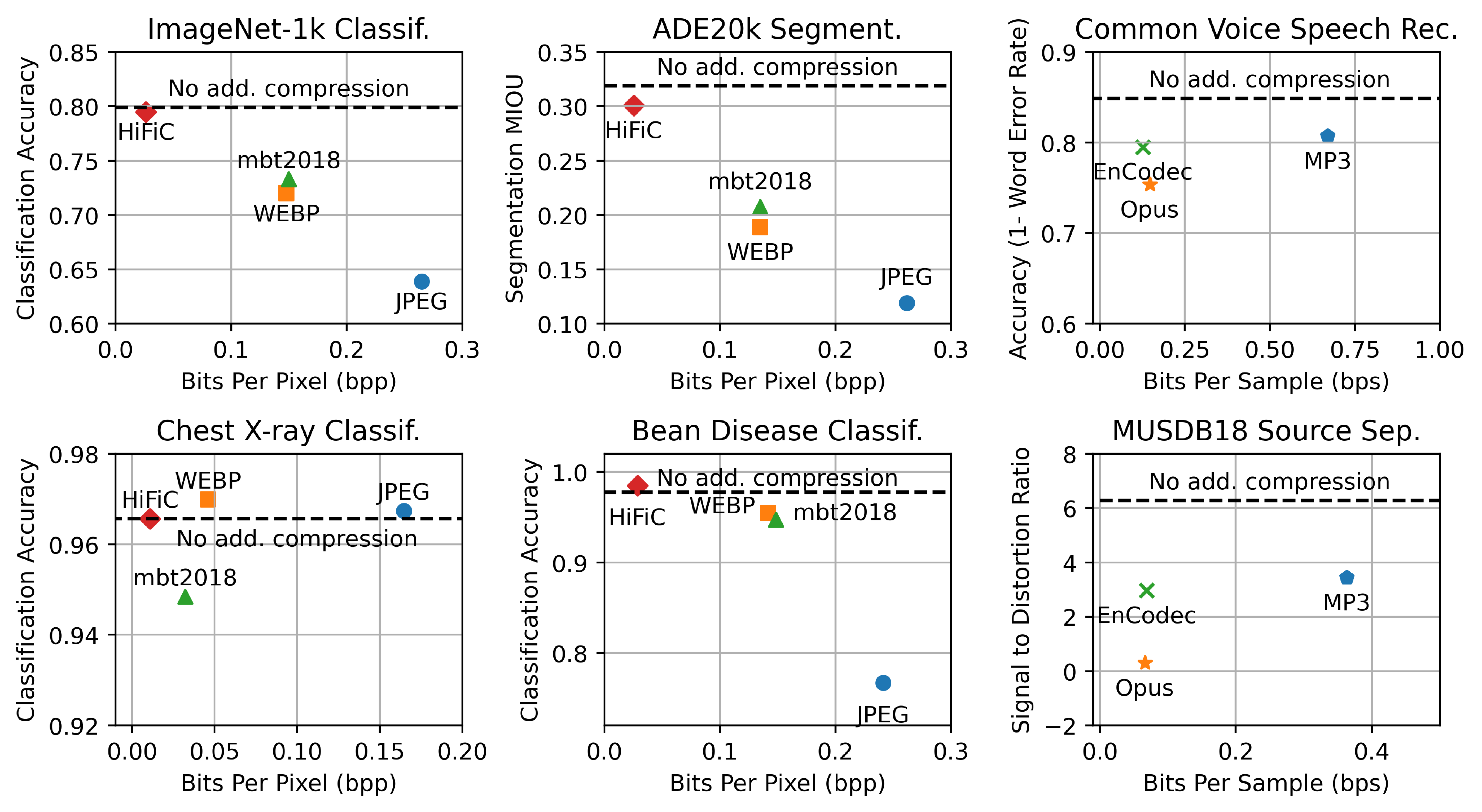}
\end{center}
\caption{\label{fig:results}%
Performance on various machine perception tasks when using different types of lossy compression.}
\end{figure}

Our evaluation across multiple datasets and machine perception tasks reveals key insights into the impact of lossy compression on machine perceptual quality. The results are shown in Figure~\ref{fig:results} and are summarized in Tables~\ref{tab:image_compression_comparison} and~\ref{tab:audio_compression_comparison}. For image-based tasks, LPIPS is a better predictor of downstream performance than PSNR, and generative compression (HiFiC) performs the best at all but one of the tasks (Chest X-ray) despite having the lowest average bitrate, and consistently achieves results close to the uncompressed baseline. In the audio domain, similar trends are observed; the audio quality measured by CDPAM is better predictor of downstream performance than PSNR, and, among the methods tested, EnCodec provides the best trade-off between rate and downstream performance for both datasets.

\begin{table}[tb]
\small
\centering
\caption{Summary of image results.}
\label{tab:image_compression_comparison}
\begin{tabular}{lcccccc}
\toprule
Metric & Dataset & Baseline & JPEG & WEBP & MBT2018 & HiFiC \\
\midrule
PSNR & ImageNet &  & 23.18 & 24.76 & \textbf{26.67} & 26.25 \\
     & ADE20k &  & 23.85 & 25.59 & \textbf{28.05} & 27.70 \\
     & Bean Disease &  & 20.76 & 21.96 & \textbf{22.92} & 21.82 \\
     & Chest X-ray &  & 30.00 & 32.69 & 34.70 & \textbf{36.44} \\
\midrule
LPIPS & ImageNet &  & 6.109 & 7.017 & 7.945 & \textbf{10.834} \\
      & ADE20k &  & 7.134 & 7.959 & 8.992 & \textbf{11.81} \\
      & Bean Disease &  & 5.716 & 6.749 & 6.899 & \textbf{9.779} \\
      & Chest X-ray &  & 6.851 & 7.584 & 7.799 & \textbf{13.24} \\
\midrule
BPP & ImageNet &  & 0.2647 & 0.1478 & 0.1499 & \textbf{0.0263} \\
    & ADE20k &  & 0.2616 & 0.1347 & 0.1347 & \textbf{0.0254} \\
    & Bean Disease &  & 0.2413 & 0.1415 & 0.1484 & \textbf{0.0286} \\
    & Chest X-ray &  & 0.1646 & 0.0459 & 0.0323 & \textbf{0.0108} \\
\midrule
Classification & ImageNet & \textbf{0.799} & 0.639 & 0.720 & 0.733 & 0.795 \\
                        Accuracy& Bean Disease & 0.9774 & 0.7669 & 0.9548 & 0.9473 & \textbf{0.9849} \\
                        & Chest X-ray & 0.9656 & 0.9673 & \textbf{0.9699} & 0.9484 & 0.9656 \\
\midrule
Segment. MIOU & ADE20k & \textbf{0.3189} & 0.1191 & 0.1886 & 0.2075 & 0.3008 \\
\bottomrule
\end{tabular}
\end{table}

\begin{table}[tb]
\small
\centering
\caption{Summary of audio results.}
\label{tab:audio_compression_comparison}
\begin{tabular}{lccccc}
\toprule
Metric & Dataset & Baseline & MP3 & OPUS & Encodec \\
\midrule
PSNR & MUSDB18 &  & \textbf{29.17} & 22.17 & 24.95 \\
     & CV &  & \textbf{33.89} & 26.70 & 29.04 \\
\midrule
CDPAM & MUSDB18 &  & 38.43 & 36.46 & \textbf{45.33} \\
      & CV &  & 37.89 & 38.27 & \textbf{46.34} \\
\midrule
BPS & MUSDB18 &  & 0.3628 & \textbf{0.06615} & 0.06871 \\
    & CV &  & 0.6696 & 0.1439 & \textbf{0.1262} \\
\midrule
SDR & MUSDB18 & \textbf{6.286} & 3.440 & 0.2986 & 2.968 \\
\midrule
WRA & CV & \textbf{0.8488} & 0.8072 & 0.7535 & 0.7950 \\
\bottomrule
\end{tabular}
\end{table}

\Section{Discussion}

\paragraph{Generative compression preserves machine perceptual quality.} One area of concern is that generative compression methods like HiFiC and EnCodec, whose adversarial training objectives allow them to discard details at the encoder and re-synthesise them at the decoder, are ill-suited for use within machine perception pipelines. However, our results indicate the contrary; despite having the highest compression ratios among the methods tested, these methods performed well across all tasks, often outperforming methods with significantly higher bitrate. Unfortunately, current generative compression methods are far from being production-ready, and rely on architectures which are difficult to train, adapt, and deploy. However, recent advancements have shown remarkable inference speedup in score-based generative models ~\cite{song2023consistency} and vastly simplified training procedures for VQVAEs ~\cite{mentzer2023finite}. By incorporating such advancements and making these methods more accessible, generative compression could enable new applications, such as satellite, maritime and aerial remote sensing systems that require very high compression ratios.

\paragraph{Correlation of machine perceptual quality with deep similarity metrics.} Deep similarity metrics like LPIPS and CDPAM are known to be highly effective at predicting human perceptual quality as measured by mean opinion score (MOS). Across the six datasets tested, our results indicate that such metrics are also strongly correlated with machine perceptual quality, despite only being trained in a supervised fashion to predict human judgments of signal distortion pairs. A promising avenue for future research would be to extend the training objectives for these metrics to include machine judgments of distortion pairs, making them even more robust.

\paragraph{Pretraining on lossy datasets.} Our experiments reveal a surprising phenomenon: for models pre-trained on lossy datasets like ImageNet, additional lossy compression at test time may have negligible impact on performance, and can sometimes behave as an enhancement. For example, the top-1 classification accuracy on the bean disease dataset is higher when compressed using HiFiC (compression ratio of 839:1) than when using the original lossless images. Even more surprising is that severe JPEG compression (see Figure ~\ref{fig:image_compression_methods}) results in an increase in pneumonia classification performance on the Chest X-ray dataset, despite having the lowest quality measured by PSNR or LPIPS. Viewing lossy compression as a type of distribution shift provides one possible explanation for this phenomenon; subtle high-frequency details that only exist in lossless images never occur in pre-training datasets like the JPEG-compressed ImageNet. Exploring pre-training with lossless data may be feasible considering the moderate compression ratios (5:1) used such datasets. The development of lossless datasets at ImageNet or larger scale could be valuable for the development of neural compression systems---for both human and machine applications.  

\paragraph{Limitations and Future Directions.} By describing the limitations of this work we hope to highlight subtopics for future research. One key limitation of this study is the exclusive use of pre-trained models under the framework of full-input machine-oriented compression. While this approach offers a practical perspective on how existing models may perform using available compression methods, it does not capture the potential advantages of models that are tailored to compressed data. We do not explore other types of machine-oriented compression, such as model-splitting ~\cite{harell2023rate}. Although most of the codecs tested allow different quality settings, we only tested settings on the low end that result in severe loss.

\Section{Conclusion}

We observe that lossy compression is underutilized in common machine learning pipelines. 
Our study reveals a surprising and promising outcome: significantly high compression rates can be achieved without excessively compromising machine perceptual quality.  
Thus, more potent lossy compression can be integrated into learning pipelines, by extending current similarity metrics and optimizing generative compression for production scenarios. 
This leads to two key advantages, (1) greater accessibility of large-scale pre-training due to reduced storage requirements and (2) better utilization of high-resolution sensor data in bandwidth-restricted systems. Future research should expand the diversity of perception tasks and compression scenarios, and consider the creation of lossless datasets to explore the effect of lossy compression during pre-training in greater depth.

\Section{References}
\bibliographystyle{IEEEbib}
\bibliography{refs}

\end{document}